\documentclass[journal]{IEEEtran}
\ifCLASSINFOpdf
   \usepackage[pdftex]{graphicx}
\else
\fi
\usepackage{amsmath}

\usepackage[switch]{lineno}

\begin{document}
%
\title{ATLAS Tile calorimeter calibration and monitoring systems}
%
%
%

\author{Marija Marjanovi\'{c}\\
on behalf of the ATLAS collaboration\\
Laboratoire de Physique de Clermont\\
Universit\'{e} Clermont-Auvergne, CNRS-IN2P3\\
Aubi\`{e}re, France\\
marija.marjanovic@cern.ch}

\markboth{Journal of \LaTeX\ Class Files}%
{Shell \MakeLowercase{\textit{et al.}}: Bare Demo of IEEEtran.cls for IEEE Journals}
%



\maketitle

\begin{abstract}
The\let\thefootnote\relax\footnotetext{Copyright 2018 CERN for the benefit of the ATLAS Collaboration. Reproduction of this article or parts of it is allowed as specified in the CC-BY-4.0 license} ATLAS Tile Calorimeter (TileCal) is the central section of the hadronic calorimeter of the ATLAS experiment at LHC. This sampling calorimeter uses steel plates as absorber and scintillating tiles as active medium. The light produced by the passage of charged particles is transmitted by wavelength shifting fibers to photo-multiplier tubes (PMTs), located in the outer part of the calorimeter. The readout is segmented into about 5000 cells, each one being read out by two PMTs in parallel. To calibrate and monitor the stability and performance of the full readout chain during the data taking, a set of calibration sub-systems is used. The TileCal calibration system comprises Cesium radioactive sources, laser, charge injection elements, and an integrator based readout system. Combined information from all systems allows to monitor and to equalize the calorimeter response at each stage of the signal evolution, from scintillation light to digitization. Calibration runs are monitored from a data quality perspective and used as a crosscheck for physics runs. Data quality in physics runs is monitored extensively and continuously. Any problems are reported and immediately investigated. The data quality efficiency achieved was 99.6\% in 2012, 100\% in 2015, 98.9\% in 2016 and 99.4\% in 2017.

Based on LHC Run-I experience, all calibration systems were improved for Run-II. TileCal performance during LHC Run-II, (2015-2017), is discussed, including calibration, stability, absolute energy scale, uniformity and time resolution. Results show that the TileCal performance is within the design requirements and has given essential contribution to reconstructed objects and physics results.
\end{abstract}


%
\IEEEpeerreviewmaketitle

\section{Introduction}

\begin{figure}[!t]
\centering
\includegraphics[width=3.5in]{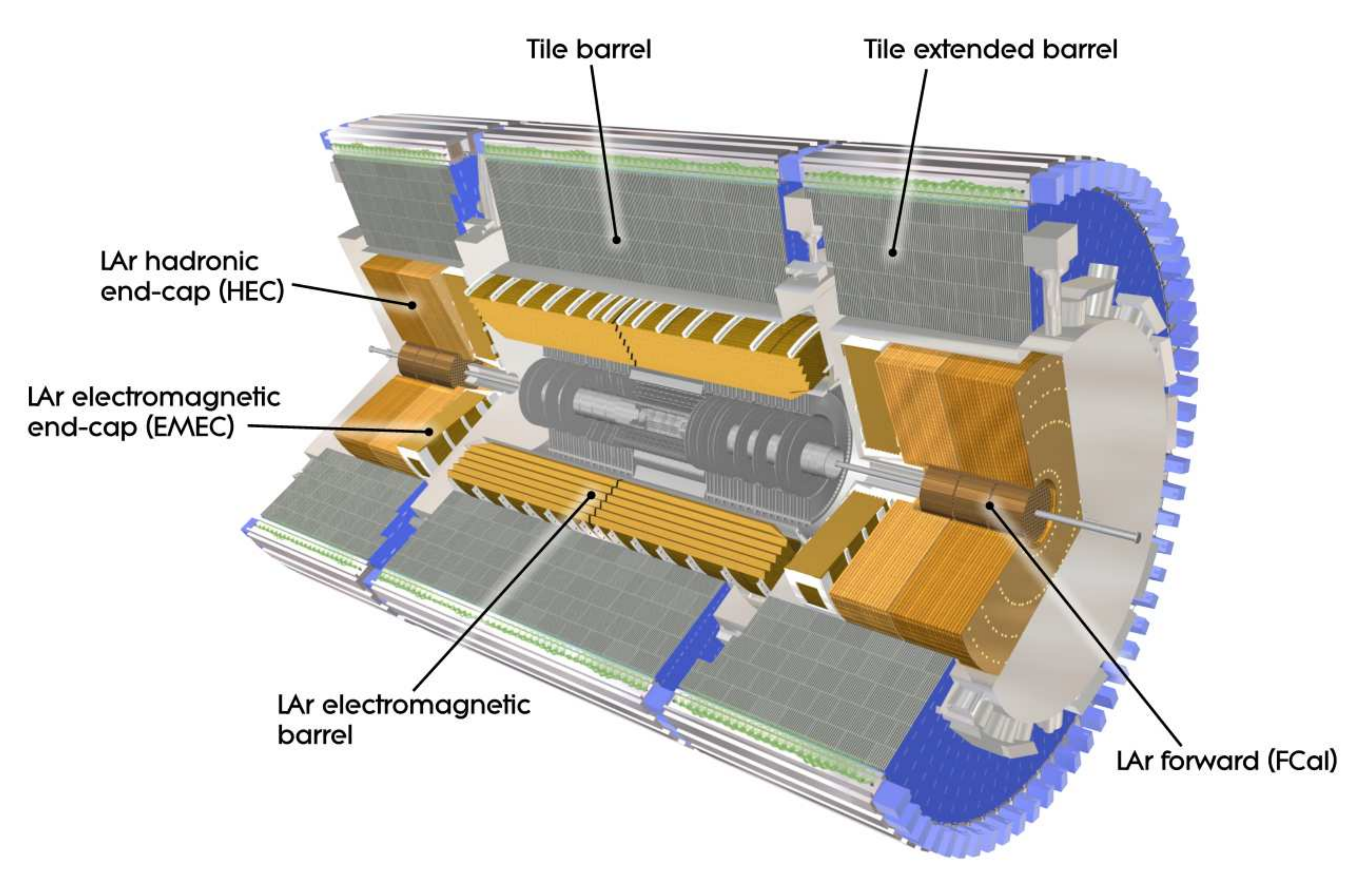}
\caption{ATLAS inner detector and calorimeters. Tile Calorimeter consists of one barrel and two extended barrel sections and surrounds the LAr barrel electromagnetic and endcap hadronic calorimeters.}
\label{fig_calorimeter}
\end{figure}

The ATLAS detector~\cite{Aad:2008zzm} is a multipurpose particle detector at the Large Hadron 
Collider (LHC) at CERN. 
It is designed to identify and measure the properties of elementary particles created in 
proton-proton ($pp$) or heavy ions collisions. It is composed of several sub-detectors, one being 
the Tile Calorimeter 
(TileCal)~\cite{Aad:2010af},~\cite{Aaboud:2018scw}, a hadronic calorimeter which allows for the 
measurements of jets, hadronically decaying 
tau leptons and missing transverse energy. TileCal is also used in the Level 1 trigger. 
Figure~\ref{fig_calorimeter} shows the electromagnetic and 
hadronic calorimeters of ATLAS. TileCal surrounds the Liquid Argon (LAr) barrel electromagnetic and endcap 
hadronic calorimeters.

\begin{figure}[!t]
\centering
\includegraphics[width=3.5in]{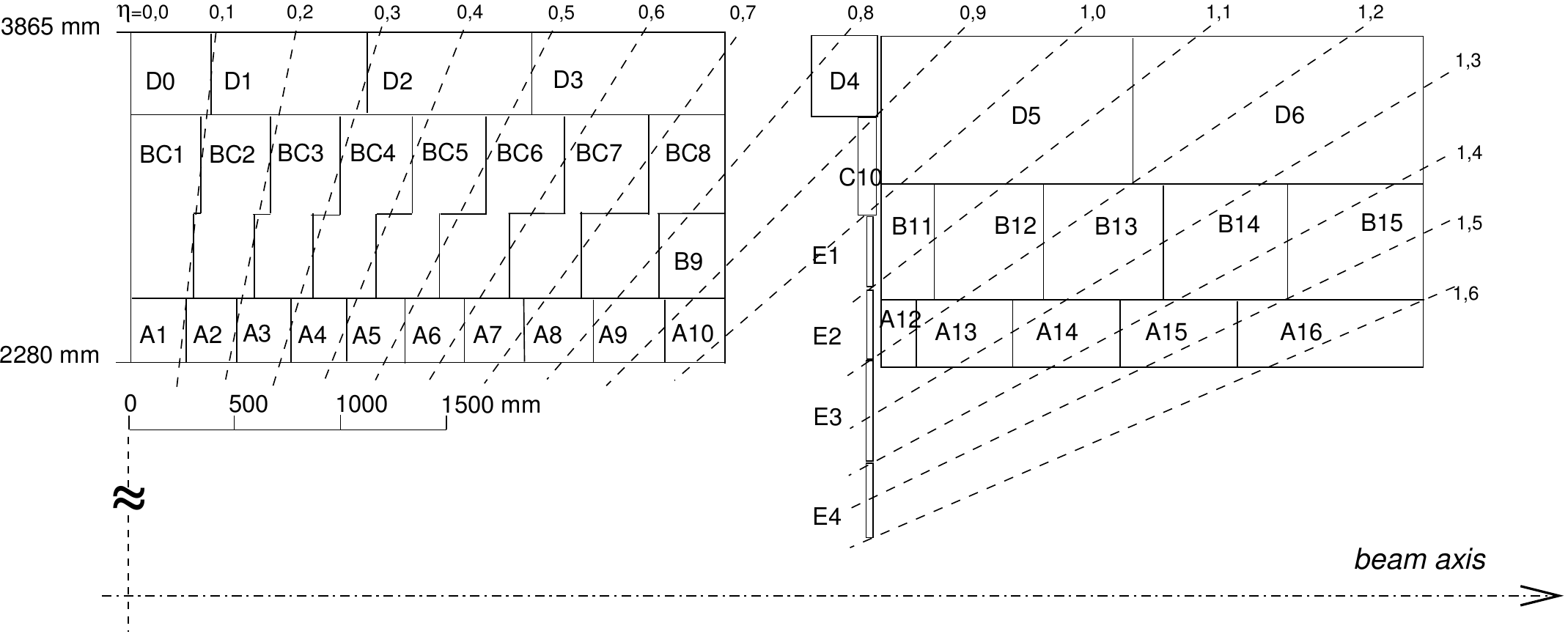}
\caption{Schematic showing the TileCal cell and scintillator structure.}
\label{fig_tilecal_schema}
\end{figure}

TileCal is a sampling calorimeter: it consists of an absorber layer made of iron plates, 
that are used to stop the particles, and a detection layer made of scintillating tiles. It has around 5000 
cells each one read by two fibers, making a total of 1000 readout channels. TileCal covers 
the central region of ATLAS ($|\eta | < 1.7$). It consists of one barrel, and two extended barrels in $\eta$, 
and 64 modules in $\phi$. Each cylinder has three longitudinal layers (A, B/C and D). 
One additional layer (E) covers gap and crack regions. Figure~\ref{fig_tilecal_schema} shows the 
TileCal cell and scintillator structure.

Light is produced in the scintillating tiles when the particle passes through it. This light is then 
converted into electric currents by photomultiplier tubes (PMTs). Signal from the PMTs is then 
shaped and amplified using 
two gains (high and low gain) with a ratio 1:64, with a 10-bit ADC by digitizers. 
Signals are sampled and digitized at 40~MHz. Events are retained if they pass Level 1 
trigger. There is also an integrator system that measures the integrated current from the PMTs.

\section{TileCal Calibration systems}

\begin{figure}[!t]
\centering
\includegraphics[width=3.5in]{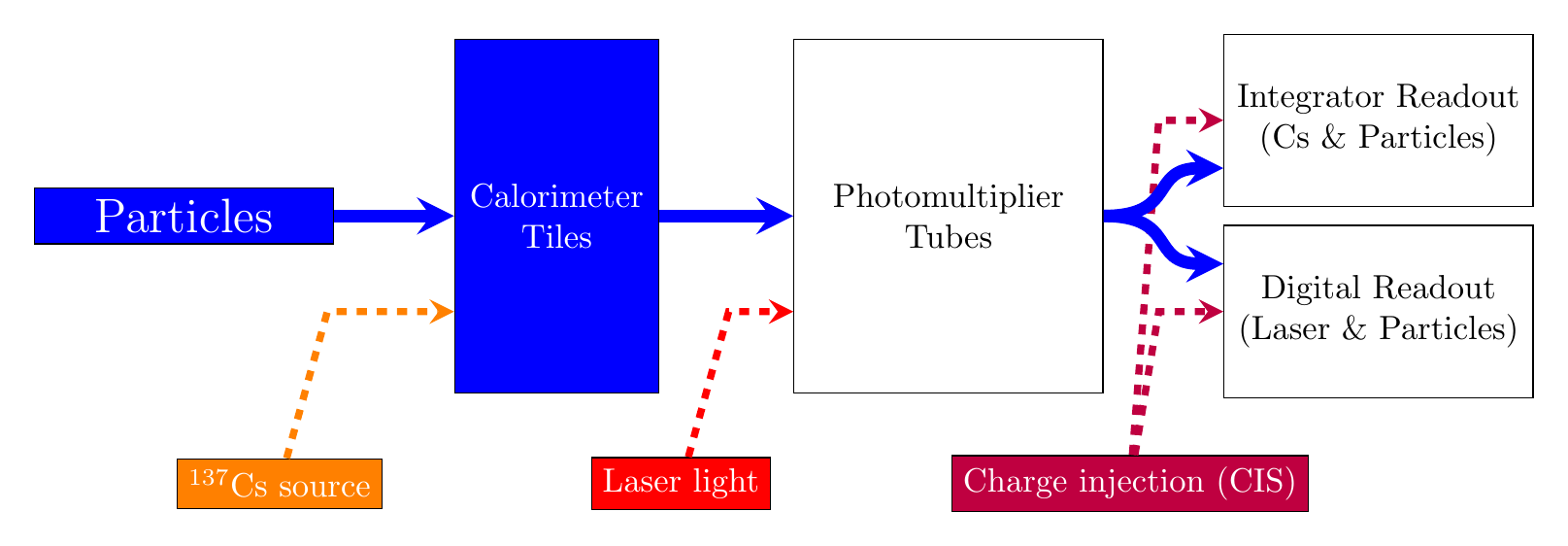}
\caption{Flow diagram of the readout signal path of different TileCal calibration systems.}
\label{fig_tilecal_calibration}
\end{figure}

The signal coming from each channel is converted into energy and that relation has to be carefully 
calibrated and monitored. There are several factors that affect it, like stability of the high voltage 
supply, the stress of the PMTs, the failures of the readout electronics and also the ageing of the 
optics. 
In order to monitor the effect of all these changes, several calibration systems are 
used~\cite{Anderson:2005ym} as shown in Figure~\ref{fig_tilecal_calibration}. 
The paths of different calibration and monitoring systems are partially overlapping, allowing for cross-checks and an easier identification of component failures.
The calibration constants are being calculated 
and applied to convert measured ADC counts into energy. 
The reconstructed energy of each channel, E(GeV), is derived from the raw response, A(ADC), as 
follows:
\begin{multline}
E(GeV) = A(ADC) \times C_{ADC \to pC} \times C_{pC \to GeV} \\
\times C_{Cesium} \times C_{Laser}
\end{multline}
where 
$C_{ADC \to pC}$ stands for conversion of the ADC counts to charge in pC. $C_{pC \to GeV}$ is 
the relation between the detector response, in pC, and the energy deposit, in GeV. 
Finally, $C_{Cesium}$ and $C_{Laser}$ are used to maintain the energy scale constant. 
While $C_{pC \to GeV}$ was fixed during dedicated test beam campaigns, the remaining calibration 
constants are provided by individual systems.

\subsection{Cesium calibration}

The calibration of optic components of TileCal and PMTs is performed using a movable Cesium 
radioactive $\gamma$ source~\cite{Starchenko:2002ju}. The source of Cesium passes through the whole calorimeter and 
emits $\gamma$ rays of 
well known energy (662 keV). It uses an integrator read-out system, a different readout system than 
the one used for physics. 
The cesium system is used for checking the quality of the optics and full system readout (scintillators, 
fibers, PMTs and electronics), and to equalize the response of all read-out channels.
It also monitors the cell electromagnetic scale in time. 

\begin{figure}[!t]
\centering
\includegraphics[width=3.5in]{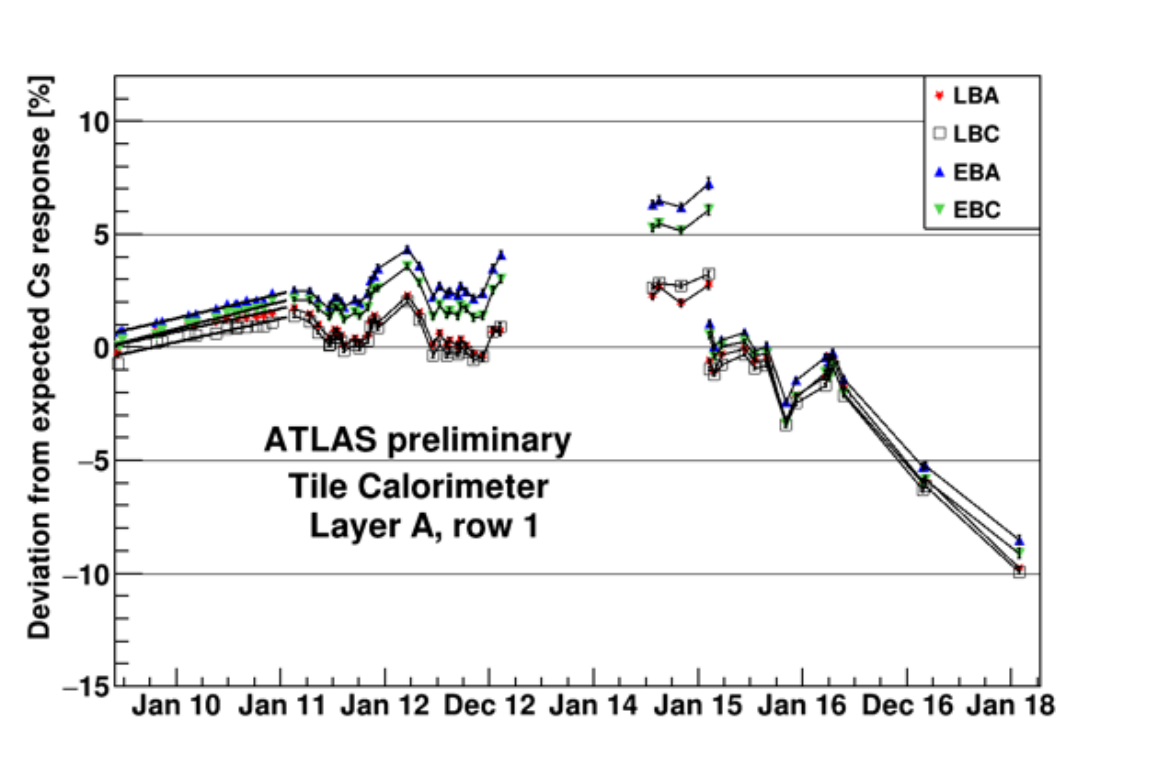}
\caption{Variation of TileCal response over a period of 8 years measured in Cesium calibration runs averaged over all cells in a given sample.}
\label{fig_cesium}
\end{figure}

The Cesium system was upgraded for the Run-II of the LHC. This included improvements of stability 
and safety (new water storage system, lower pressure and precise water level metering).
The precision of single channel response is now better than 0.3\%.
Figure~\ref{fig_cesium} shows the variation of TileCal response over a period of 8 years measured in 
Cesium calibration runs averaged over all cells in a given sample \cite{AppPlotsTile}. 

\begin{figure}[!t]
\centering
\includegraphics[width=3.5in]{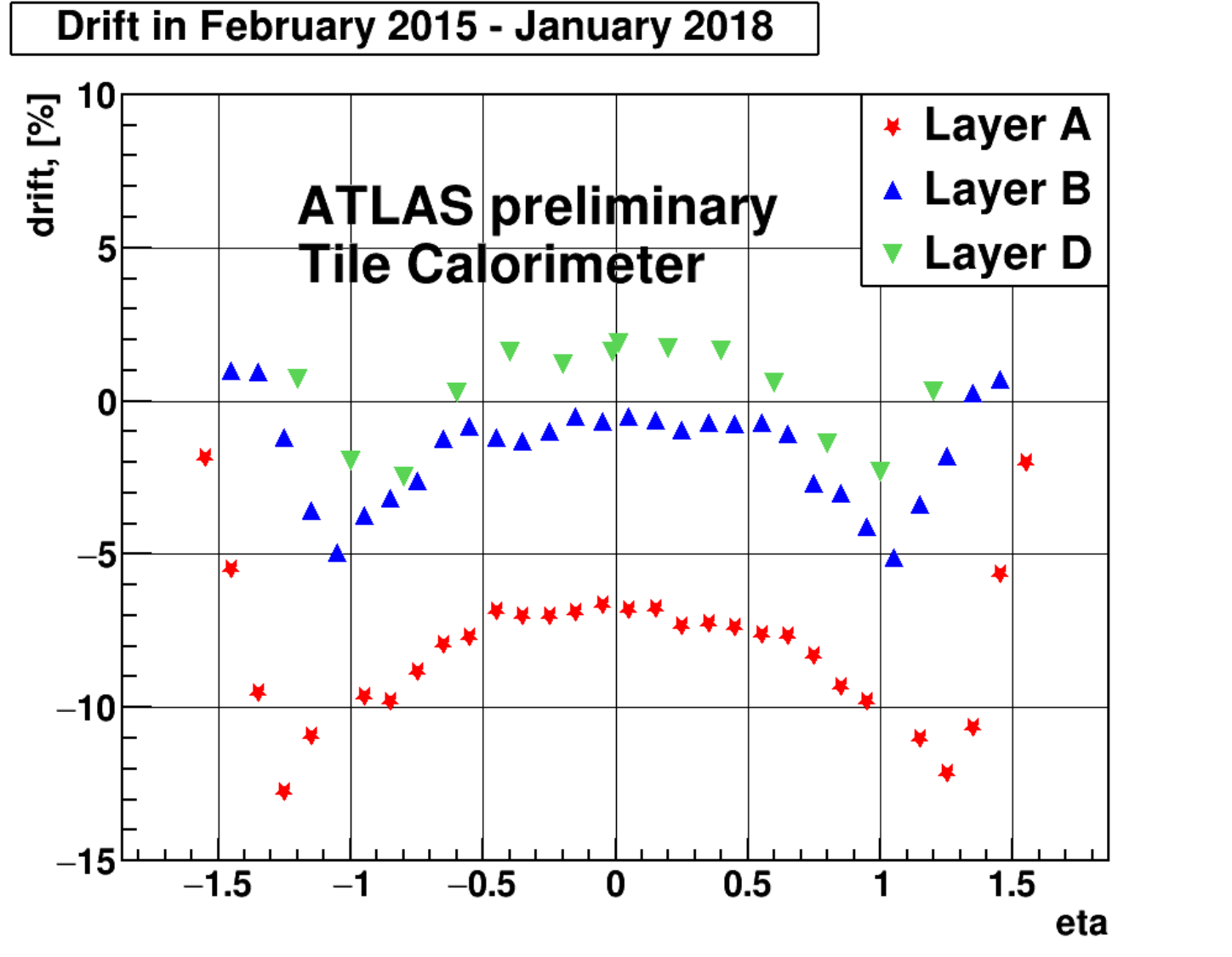}
\caption{Change of the signal in TileCal cells between February 2015 and January 2018 as a function of $\eta$ in three longitudinal samplings.}
\label{fig_cesium_2}
\end{figure}

Figure~\ref{fig_cesium_2} presents the change of the signal in TileCal cells between February 2015 
and January 2018 as a function of $\eta$ in the three longitudinal samplings. Signal is measured in 
Cesium calibration runs. Every point represents average of 64 cells (over $\phi$). The largest drift is
observed for A-cells.

\subsection{Laser calibration}

The calibration of PMT gains is performed twice a week using a laser calibration 
system~\cite{system:2016tae} during dedicated runs. 
The laser system measures the drift seen in PMT response with respect to the last cesium scan. 
It sends light pulses to the PMTs with a wavelength close to the one of physical signals (532~nm) in the absence of collisions.
Controlled amount of light is sent to the PMTs through $\sim$400 fibers.
The mean gain variation of the 
9852 TileCal channels is computed cell by cell. For each cell, the gain variation is defined as the 
mean of the gaussian function that fits the gain variation distribution of the channels associated to 
this cell. A total of 64 modules in $\phi$ are used for each cell while known pathological channels 
are 
excluded. The observed down-drift mostly affects cells at inner radius, that are the cells with higher 
current. Maximal drift is observed in E and A cells which are the cells with the highest energy 
deposits.

\begin{figure}[!t]
\centering
\includegraphics[width=3.5in]{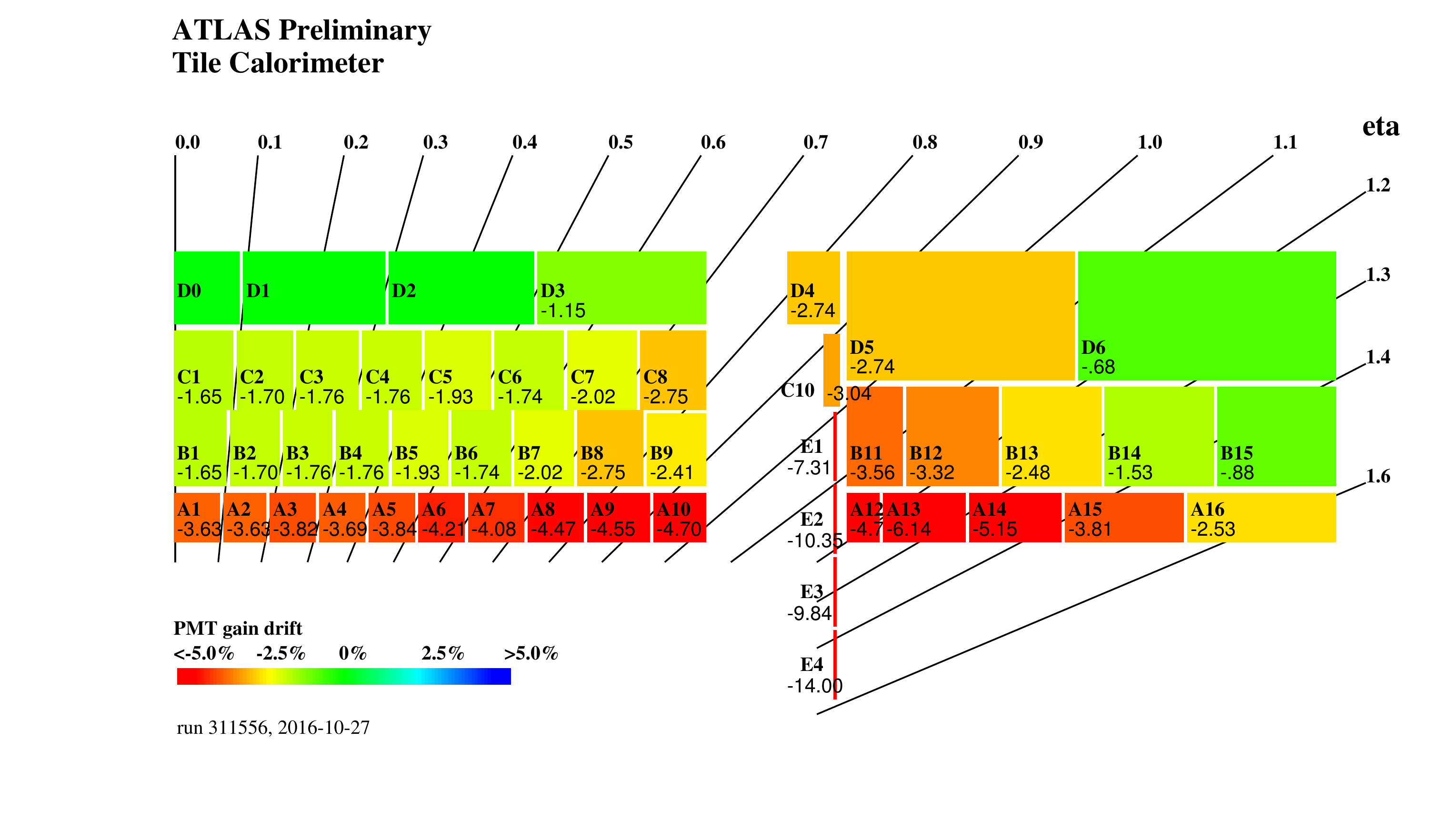}
\caption{The mean gain variation (in \%) in the ATLAS TileCal PMTs that read 
the signal deposited in each channel, as a function of $\eta$ and radius, between the 24 May 2016 
and the 27 October 2016 (given as an example).}
\label{fig_laser_eta}
\end{figure}

Figure~\ref{fig_laser_eta} shows the mean gain variation (in \%) in the ATLAS TileCal PMTs that 
read 
the signal deposited in each channel, as a function of $\eta$ and radius, between the 24 May 2016 
and the 27 October 2016 (given as an example).

\begin{figure}[!t]
\centering
\includegraphics[width=3.5in]{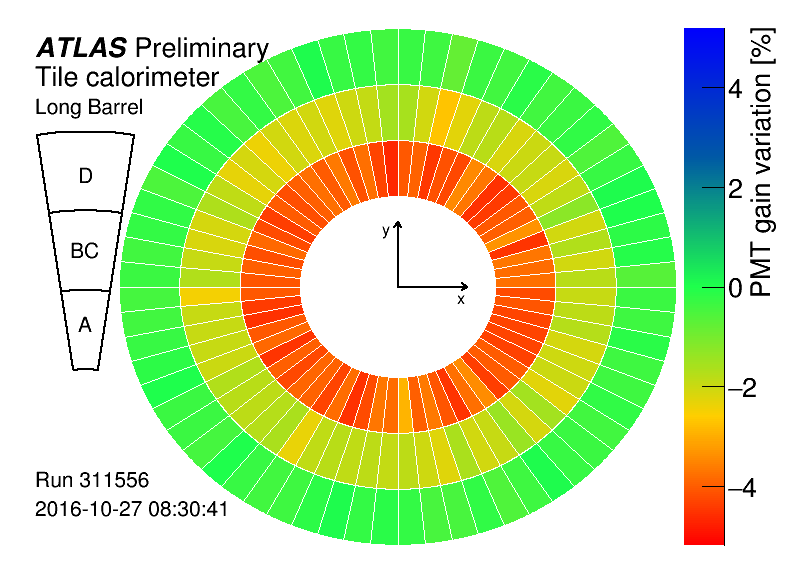}
\caption{The mean gain variation (in \%) in the ATLAS TileCal PMTs that read 
the signal deposited in each channel, as a function of $\phi$ and radius, between the 24 May 2016 
and the 27 October 2016 (given as an example).}
\label{fig_laser_phi}
\end{figure}

Figure~\ref{fig_laser_phi} depicts the mean gain variation (in \%) of the channels of the ATLAS Tile calorimeter as a function of the polar angle ($\phi$) and the layer. 
The PMT mean gain variation is defined as the mean of the gaussian function that fits the gain 
variation distribution of the channels, arranged into $\phi$ bins. One $\phi$ bin corresponds to one 
module per partition. 
The channels with known 
problems are not taken into account.

The laser system underwent an upgrade for Run-II of the LHC. 
The electronics and optical components were improved, which led to better control of the emitted 
light.

\subsection{Charge injection system}

Charge Injection System (CIS) injects a signal of known charge and measures the electronic 
response. 
It checks the full ADC range (0-800~pC) and has two gains for each PMT (low gain and high gain). 
CIS calibration is taken at least twice a week and it measures the pC/ADC conversion factor and 
corrects for non-linearities in low gain. It is also used to calibrate analog Level 1 trigger.

\begin{figure}[!t]
\centering
\includegraphics[width=3.5in]{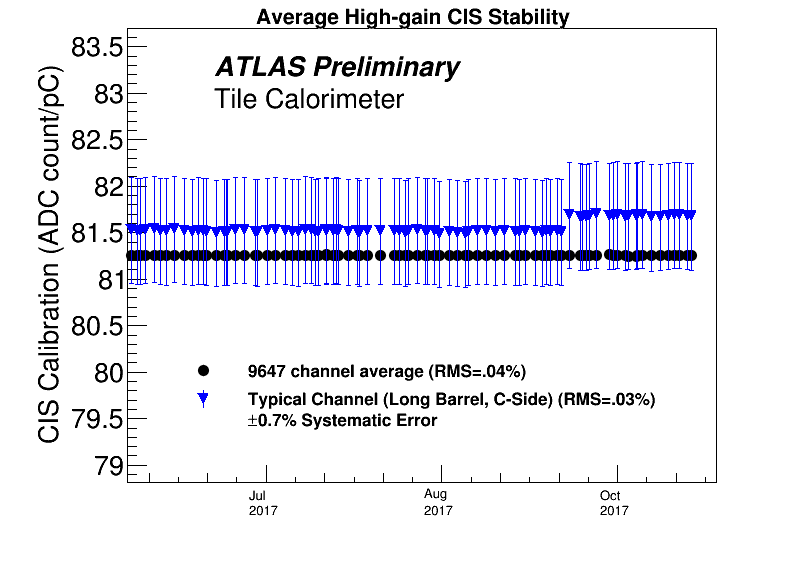}
\caption{CIS calibration constant averages of all the high gain ADC's for each CIS calibration run 
from 15 May 2017 to 05 December 2017, plotted as black 
circles. The CIS constants from a typical channel (LBC19, Channel 13) are additionally plotted as 
blue triangles for comparison.}
\label{fig_cis_hg}
\end{figure}

Figure~\ref{fig_cis_hg} shows the detector-wide CIS calibration constant averages of all the high 
gain ADC's for each CIS calibration run from 15 May 2017 to 05 December 2017 (given as an example), plotted as black 
circles. The CIS constants from a typical channel (LBC19, Channel 13) are additionally plotted as 
blue triangles for comparison. The RMS values on the plot are indicative of the fluctuation present 
in calibrations. In addition, there is a 0.7\% systematic uncertainty present in individual calibrations, 
represented by the blue error bars. Problematic channels are not included when calculating the 
mean.

The CIS measurements have a precision of 0.7\% and the average of all channels doesn't 
change by more than 0.03\%.

\subsection{Integrator system}

The integrator system~\cite{Fracchia:1710463} integrates PMT signals over a large time window ($\sim10\mu s$).
It monitors and measures the response to the source during Cesium scans, and during physics 
runs, it measures the detector response to the minimum bias events. The minimum bias events 
are soft parton interactions that dominate the high energy $pp$ collisions. 
The integrator system also provides an additional way to monitor the instantaneous luminosity in 
ATLAS as it allows for an independent measurement given an initial calibration (luminosity 
coefficient).
This is possible as the measured currents are linearly dependent on the instantaneous luminosity. 
The response stability is used to produce calibration constants in absence of Cesium calibration
and to calibrate E-cells and MBTS (Minimum Bias Trigger Scintillator cells), which are not 
instrumented by Cesium system. 
The integrator system monitors the full optical chain.

\begin{figure}[!t]
\centering
\includegraphics[width=3.5in]{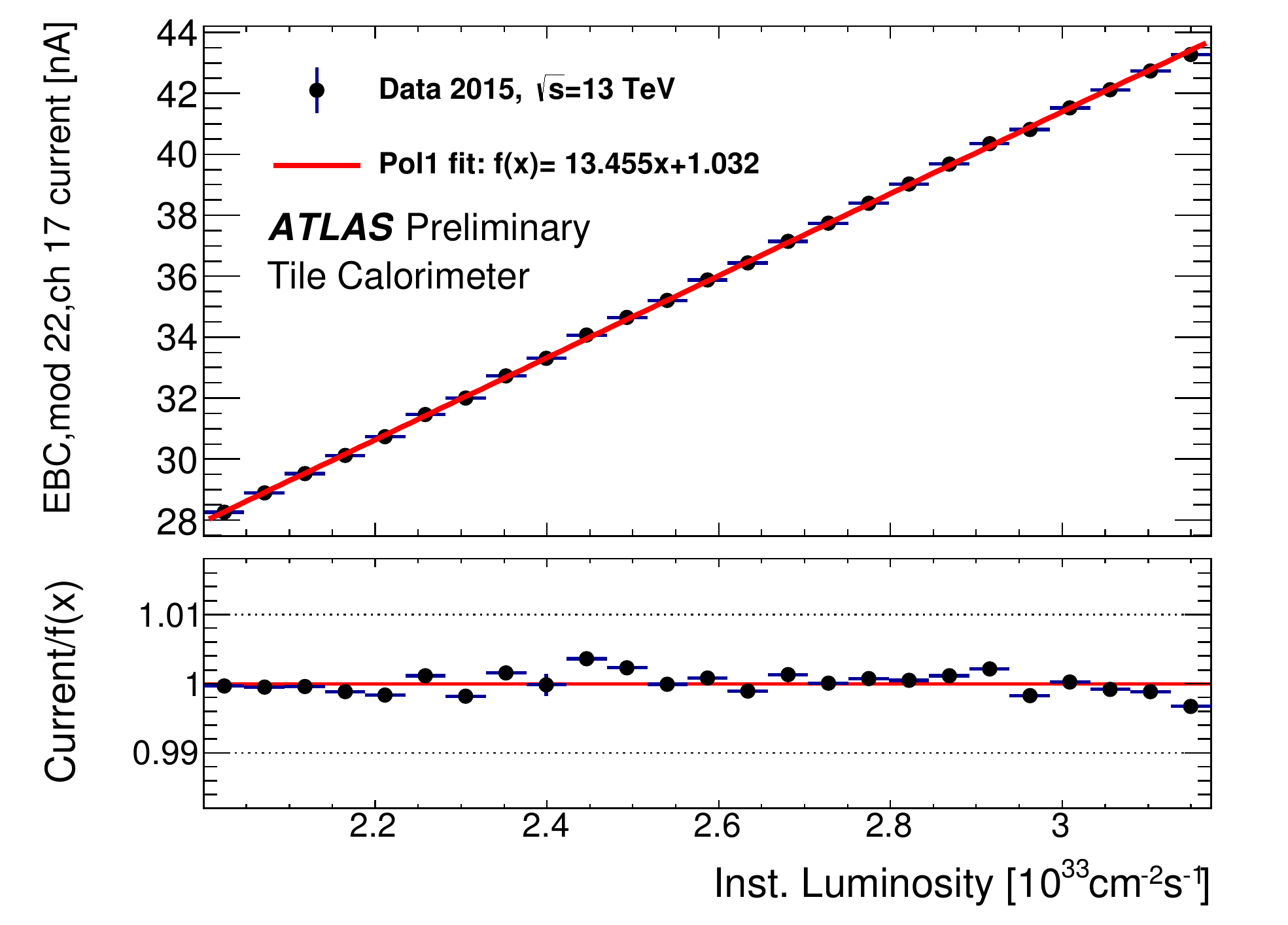}
\caption{Average current (measured by PMT 17 of module 22 
in the C-side Extended Barrel) vs. instantaneous luminosity measured by the integrator system 
during the $pp$ collision data taking in 2015. The dashed region presents the integrated 
luminosity delivered to the ATLAS detector.}
\label{fig_current_vs_lumi}
\end{figure}

Figure~\ref{fig_current_vs_lumi}
presents the average current vs. instantaneous luminosity measured by the integrator system 
during the $pp$ collision data taking in 2015. The currents were measured by PMT 17 of module 22 
in the C-side Extended Barrel. This channel belongs to cell D5, which is stable within 1-2\% over 
the 2015 data taking period. The considered runs are spread over the whole data taking period and 
fulfil the quality requirements used for physics analyses. A linear fit has been performed. The fit 
parameters are given in the upper panel of the plot. The lower panel shows the ratio of the current 
over the fitted function. A good linear description of the current vs instantaneous luminosity can be 
observed. 
The stability of the integrator gains for each channel is better than 0.05\% and the average stability is 
better than 0.01\%.

\begin{figure}[!t]
\centering
\includegraphics[width=3.5in]{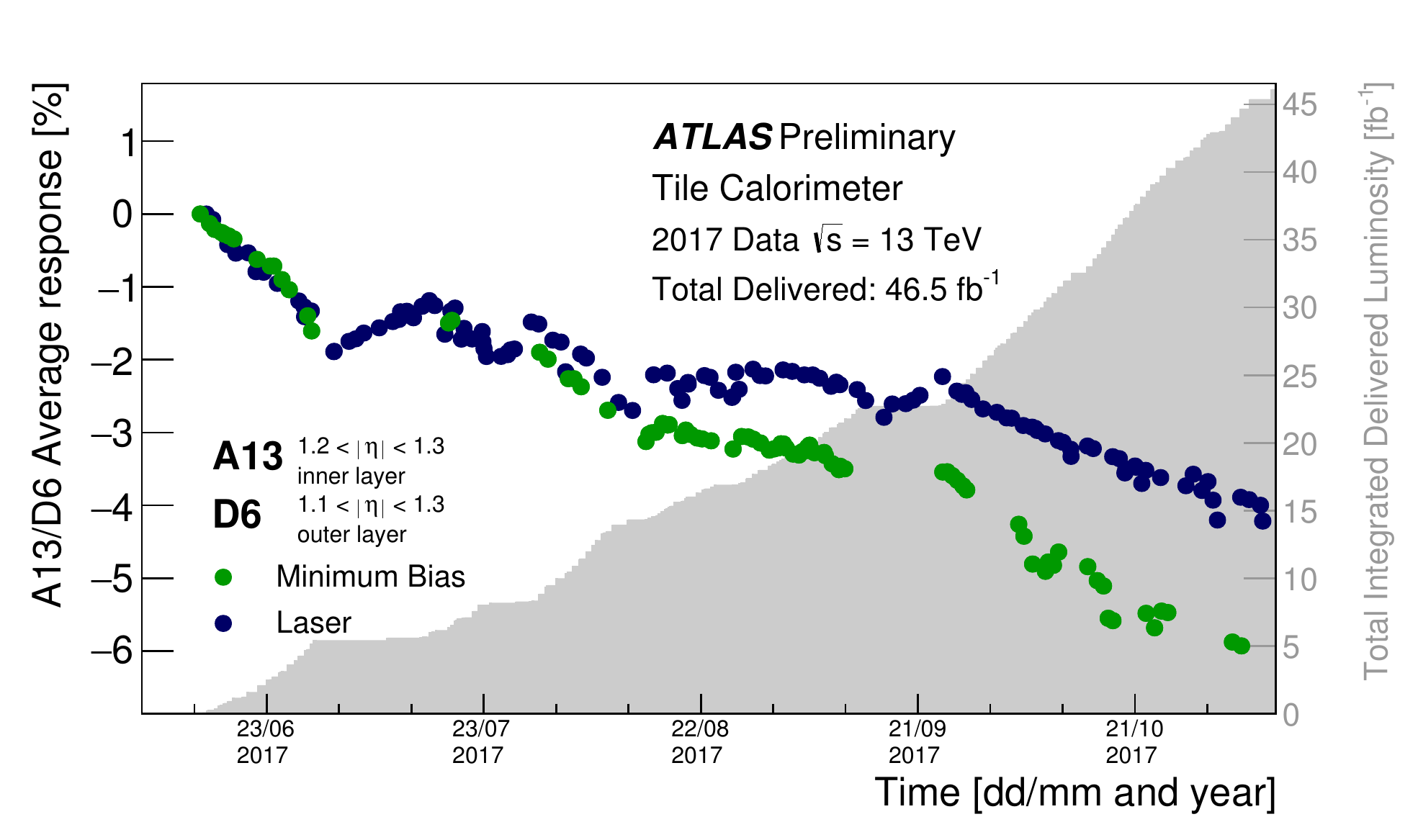}
\caption{Variation of the response to integrator (labeled Minimum Bias) and laser systems for 
cell A13 (one of the most 
irradiated cells) in the inner layer of the Extended Barrel, covering the region $1.2 < |\eta| < 1.3$, 
as a function of time.}
\label{fig_integrator}
\end{figure}

The laser and integrator systems are used in parallel to monitor TileCal. 
The integrator system monitors PMT response drift and scintillator ageing, 
while the laser system monitors only PMT response drifts.
Figure~\ref{fig_integrator} shows the variation of the response to integrator (labeled Minimum Bias) and laser systems for 
cell 
A13 in the inner layer of the Extended Barrel, covering the region $1.2 < |\eta| < 1.3$, as a function 
of 
time. The response variation is derived with respect to a reference cell D6 ($1.1 < |\eta| < 1.3$), 
which has exhibited less than 1\% drift throughout the 2017 collision period. Each Minimum Bias 
point represents the average of the response variation of a subset of A13 channels (one of the most 
irradiated cells), corresponding 
to the central 80\% of the total distribution. Each laser point represents the average of all channels. 
Integrator and laser data cover the period from the beginning of June to halfway November 2017. 
The response variation versus time measured by the integrator system has been normalized to the 
response variation measured by the laser system on June 12th, corresponding to the first point in 
the plot. The integrated luminosity is the total delivered during the proton-proton collision period of 
2017. As already observed in previous years the down-drifts of the PMT gains (seen by laser) 
coincide with the collision periods, while up-drifts are observed during machine development 
periods. 
The same response could be measured by Cesium, but the integrator measurement is used, 
as it is possible to do it with better time resolution and throughout the year. 
In 2016 and 2017 a systematic difference is observed between the laser and integrator 
systems and it is attributed to scintillator ageing due to irradiations. Therefore an extra calibration 
from the
integrator system is applied on some channels.

\begin{figure}[!t]
\centering
\includegraphics[width=3.5in]{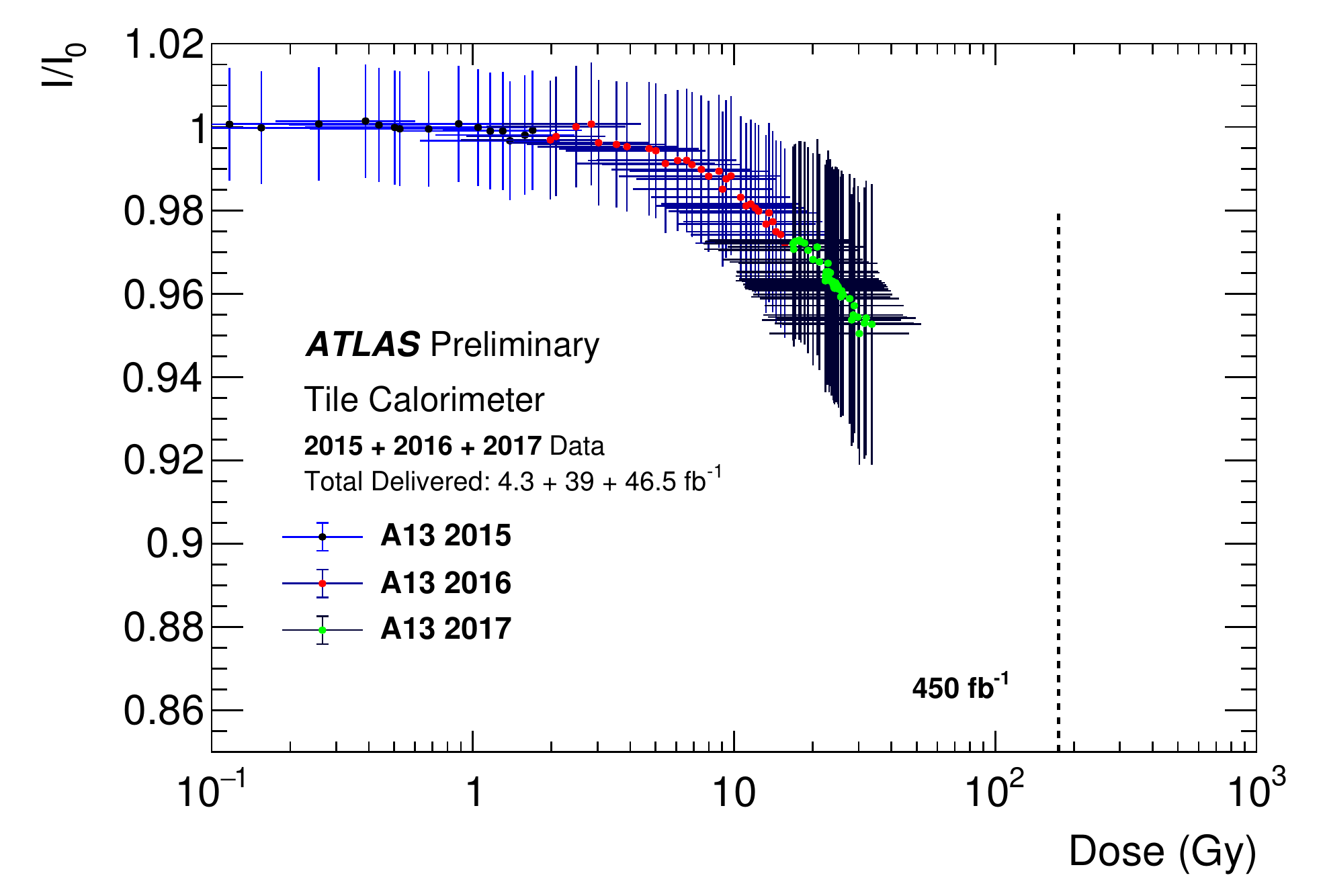} 
\caption{Relative light yield $I/I_0$ of scintillators and wavelength-
shifting fibres for the A13 cell as a function of the deposited dose.}
\label{fig_light_loss}
\end{figure}

Figure~\ref{fig_light_loss} shows the relative light yield $I/I_0$ of scintillators and wavelength-
shifting fibres for the A13 cell as a function of the deposited dose for the year 2015-17. The relative 
light yield is derived from the difference in the response to minimum bias ($\Delta R_{MB}$) events 
and laser pulses ($\Delta R_{Las}$) and interpreted as the light yield loss in scintillators and fibres 
due to irradiation, and it is defined as $I/I_0=1+(\Delta R_{MB}-\Delta R_{Las})/100\%$. The 
response variation is derived with respect to a reference cell D6. Minimum bias and laser data cover the 2015-17 
$pp$ collision period, and the integrated luminosity is the total delivered during the period. Vertical 
error bars represent the statistical and systematic uncertainties of the measurement. The horizontal 
bars represent the RMS of the different dose values within the cell volume and do not include 
systematics. For the nominal dose value the average is taken. The black vertical line represents the expected dose by the end of Run-III (450 fb$^{-1}$).

\subsection{Time calibration}

A precise time calibration is crucial as 
it is needed to 
reconstruct the energy recorded in each cell. It is used to set the phase so that a particle that is 
travelling from the collision point at the light speed would give a signal whose measured time would 
be equal to zero. The timing information is also used in some physics analysis, for example those 
that search for R-hadrons, using time of flight. Time calibration is calculated using jets and is monitored during data 
taking using laser system. Laser pulses are sent to TileCal during empty bunch crossing (1-2~Hz frequency) to calibrate timing.

\begin{figure}[!t]
\centering
\includegraphics[width=3.5in]{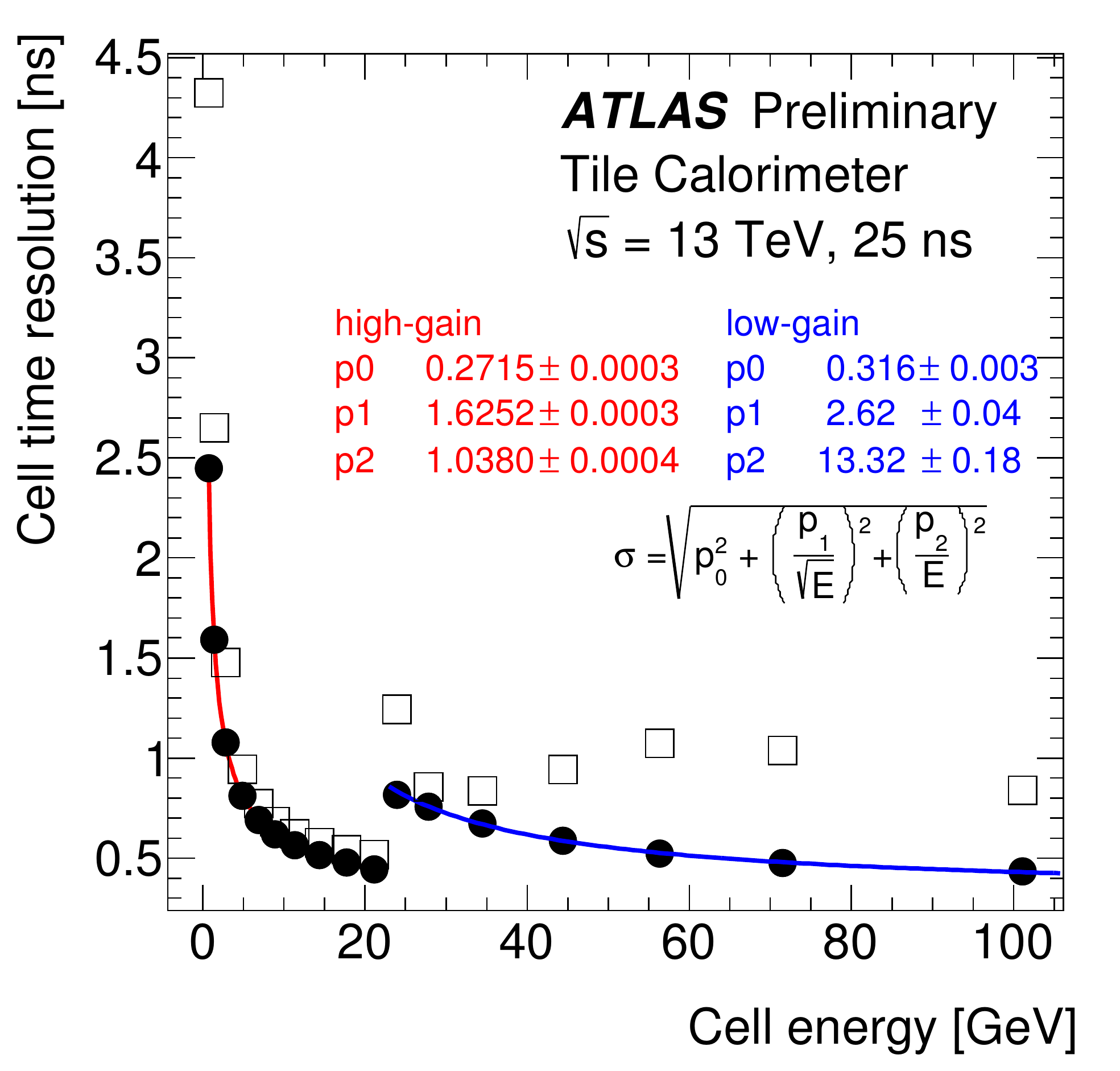}
\caption{Cell time resolution in jet events ($pp$ data at $\sqrt{s}$=13 TeV, 25 ns bunch spacing) as a function of the energy deposited in Long Barrel cells.}
\label{fig_cell_time_resolution}
\end{figure}

Figure~\ref{fig_cell_time_resolution} shows the cell time resolution in jet events ($pp$ collision data 
at $\sqrt{s}$=13~TeV with 25~ns bunch spacing) as a function of the energy deposited in Long Barrel 
cells. Run-to-run differences are accounted for. All cells belonging to reconstructed jets with 
$p_T>$20~GeV are considered, after the usual event and jet cleaning procedures are applied.
The closed circles correspond to Gaussian $\sigma$, the open squares indicate the RMS of the 
underlying 
time distributions. The resolution in Long Barrel is slightly better than in Extended Barrel cells (even 
if E-cells are excluded), due to physically smaller cells in Long Barrel compared to Extended Barrel.

\section{Performance of the TileCal}

\subsection{Noise}

The control of the noise also plays an important role in physics measurements .The total noise in 
each cell of the calorimeter comes 
from two sources: electronic noise and noise coming from the pile-up. Electronic noise is controlled 
by measurements taken in the dedicated runs when there is no signal in the detector. Pile up 
contribution to noise comes from the multiple interactions that take place at the same bunch 
crossing, or from the events from the previous/ following bunch crossings.
Electronic noise stays at the level below 20~MeV for most of the cells. 
Noise is measured regularly with calibration runs. 
In 2014 new power supplies were installed giving better electronics stability and more gaussian 
noise shape.

\begin{figure}[!t]
\centering
\includegraphics[width=3.5in]{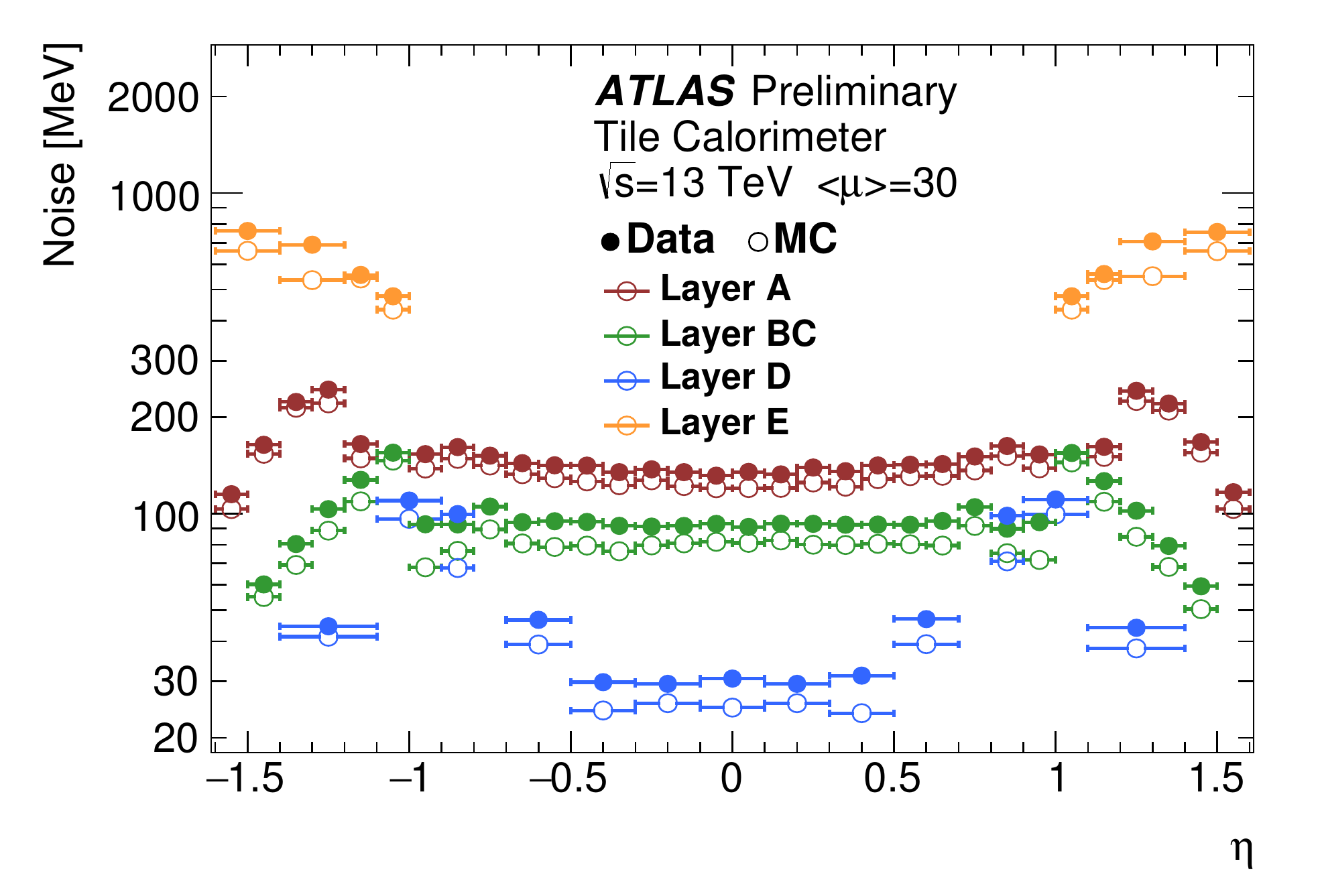}
\caption{Noise distributions in different TileCal cells, represented as a function of $\eta$.}
\label{fig_pileup_eta}
\end{figure}

Figure~\ref{fig_pileup_eta} shows the noise distributions in different TileCal cells as a 
function of $\eta$ for zero-bias data collected in 2016 (runs 310247, 310249, 310341, 310370) and 
Pythia 8 Monte Carlo simulation with A3 Minimum Bias tune at a center-of-mass energy of 13~TeV 
with a bunch spacing of 25~ns selected with an average number of interactions  per bunch crossing 
of 30. The noise was estimated as the standard deviation of the measured cell energy distribution. 
The data (Monte Carlo) is represented with closed (open) markers. The cells of Layer A, Layer BC, 
Layer D and Layer E are shown with different colors. 

\begin{figure}[!t]
\centering
\includegraphics[width=3.5in]{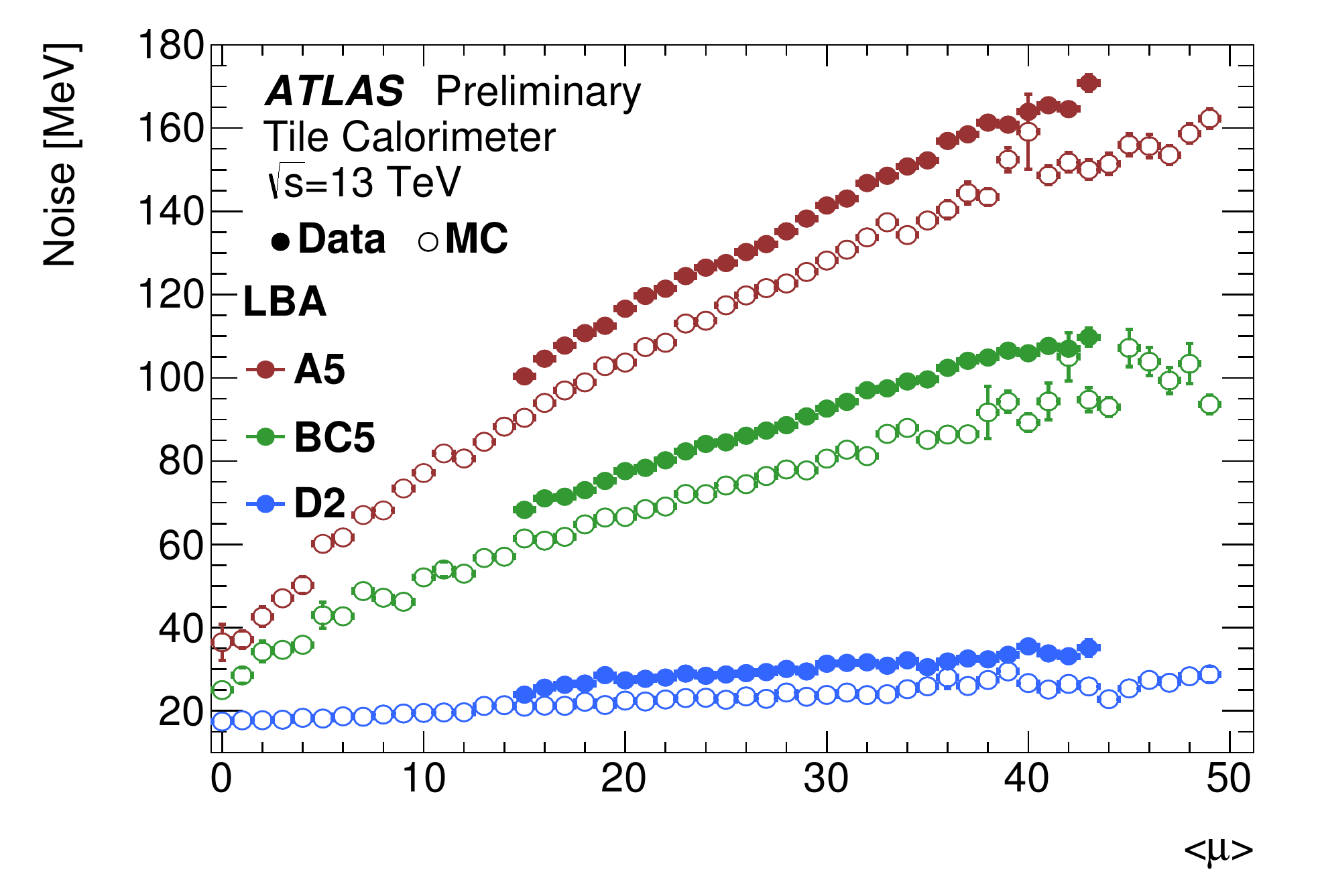}
\caption{Noise dependence with the average number of 
interactions per bunch crossing, $\langle \mu\rangle$.}
\label{fig_pileup_mu}
\end{figure}

Figure~\ref{fig_pileup_mu} presents the TileCal noise dependence with the average number of 
interactions per bunch crossing, $\langle \mu\rangle$, for several zero-bias data runs of 2016 
with 25~ns bunch spacing at a center-of-mass energy of 13~TeV (runs 310247, 310249, 
310341, 310370) and Pythia 8 Monte Carlo simulation. 
The noise was estimated as the standard deviation of the energy distribution per cell. The data 
(Monte Carlo) is represented with closed (open) markers. The noise is shown for one TileCal 
tower with cells A5, BC5 and D2 in $0.4 < \eta < 0.5$. Only $\langle \mu\rangle > 15$ are shown in 
data were the reliable statistics is observed. 


\subsection{Data quality}

\begin{figure}[!t]
\centering
\includegraphics[width=3.5in]{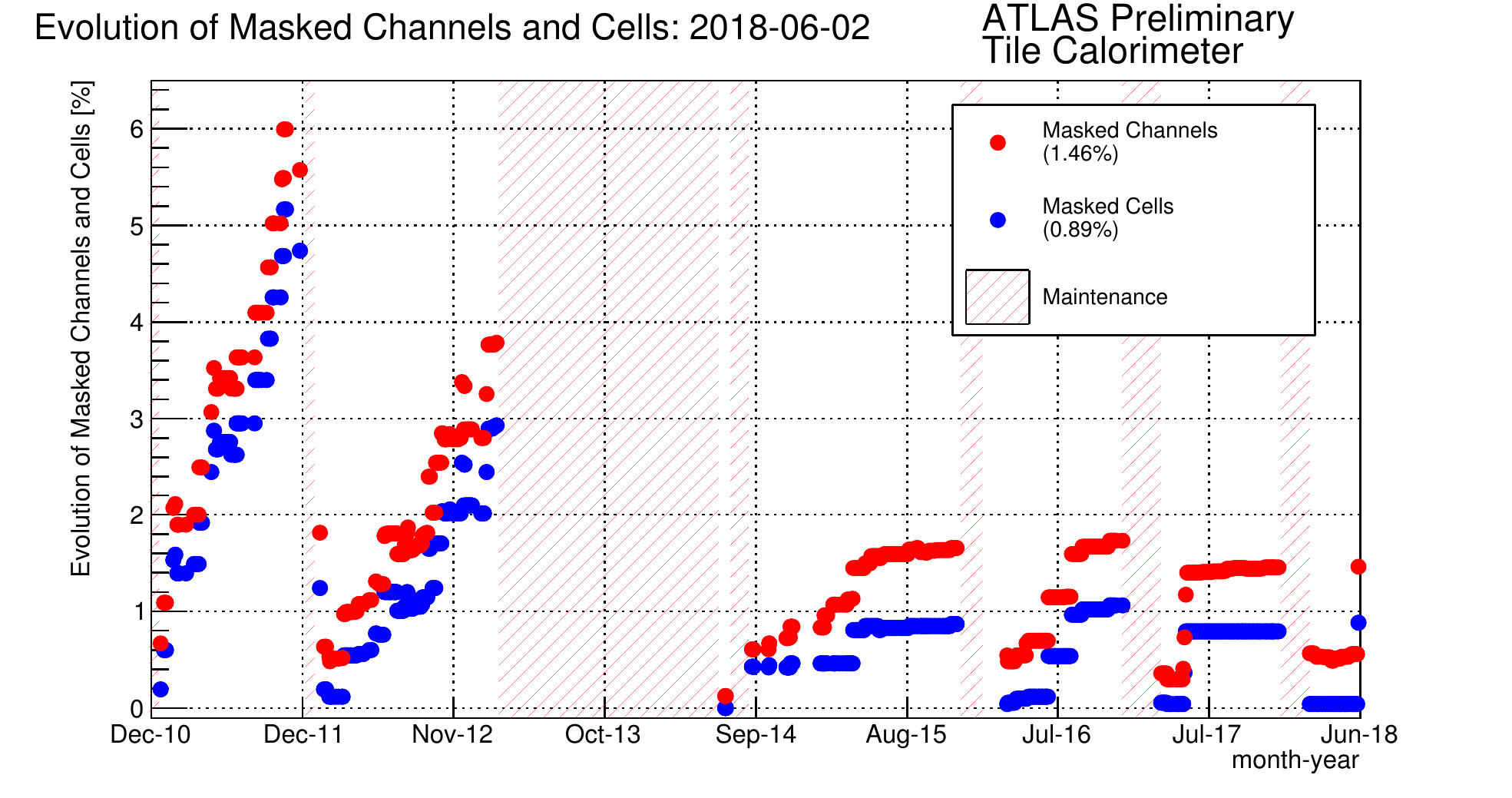}
\caption{Evolution of Cell Masking in TileCalorimeter.}
\label{fig_cell_masking}
\end{figure}

TileCal channels are monitored for hardware issues, timing offsets and miscalibrations. 
Problematic channels are 
identified and masked. During maintenance periods all identified issues are fixed. 
Data Quality efficiency was 100\% in 2015, 98.8\% in 2016 and  99.4\% in 2017.
Figure~\ref{fig_cell_masking} presents the evolution of cell masking in TileCal. It shows the 
percentage of all cells and channels in the detector that are masked as a function of time starting 
from December 2010. The recent detector status corresponding to the 2nd June, 2018 is noted. The 
hatched area represents the maintenance period of the detector. The legend includes the amount of 
masked cells (0.89\%) and masked channels (1.46\%) at that time.

\begin{figure}[!t]
\centering
\includegraphics[width=3.5in]{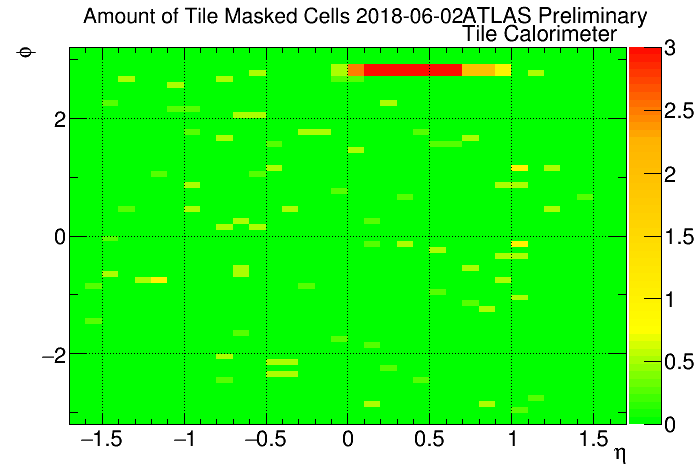}
\caption{Representation of the number of TileCal masked cells in $\eta -\phi$ plane.}
\label{fig_cell_masking_eta_phi}
\end{figure}

Figure~\ref{fig_cell_masking_eta_phi} 
shows the number of cells masked in each $\eta -\phi$ bin. The bin is chosen to be $0.1 \times 0.1$, corresponding to the calorimeter tower definition. TileCal cells can have one or two readout 
channels. The analog signal from each channel is then amplified in separate high (HG) and low (LG) 
branches and digitized by two ADCs. The bin color corresponding to integer numbers is the 
integrated number of cells in calorimeter tower. The non-integer numbers mean the integrated 
fraction of masked ADCs in tower.

\subsection{Single particle response}

An important characteristic of the TileCal is the ratio of energy at electromagnetic scale to track 
momentum $\langle E/p\rangle$ for isolated charged hadrons in minimum bias events. 
It is used to evaluate calorimeter uniformity and linearity during data taking. 

\begin{figure}[!t]
\centering
\includegraphics[width=3.5in]{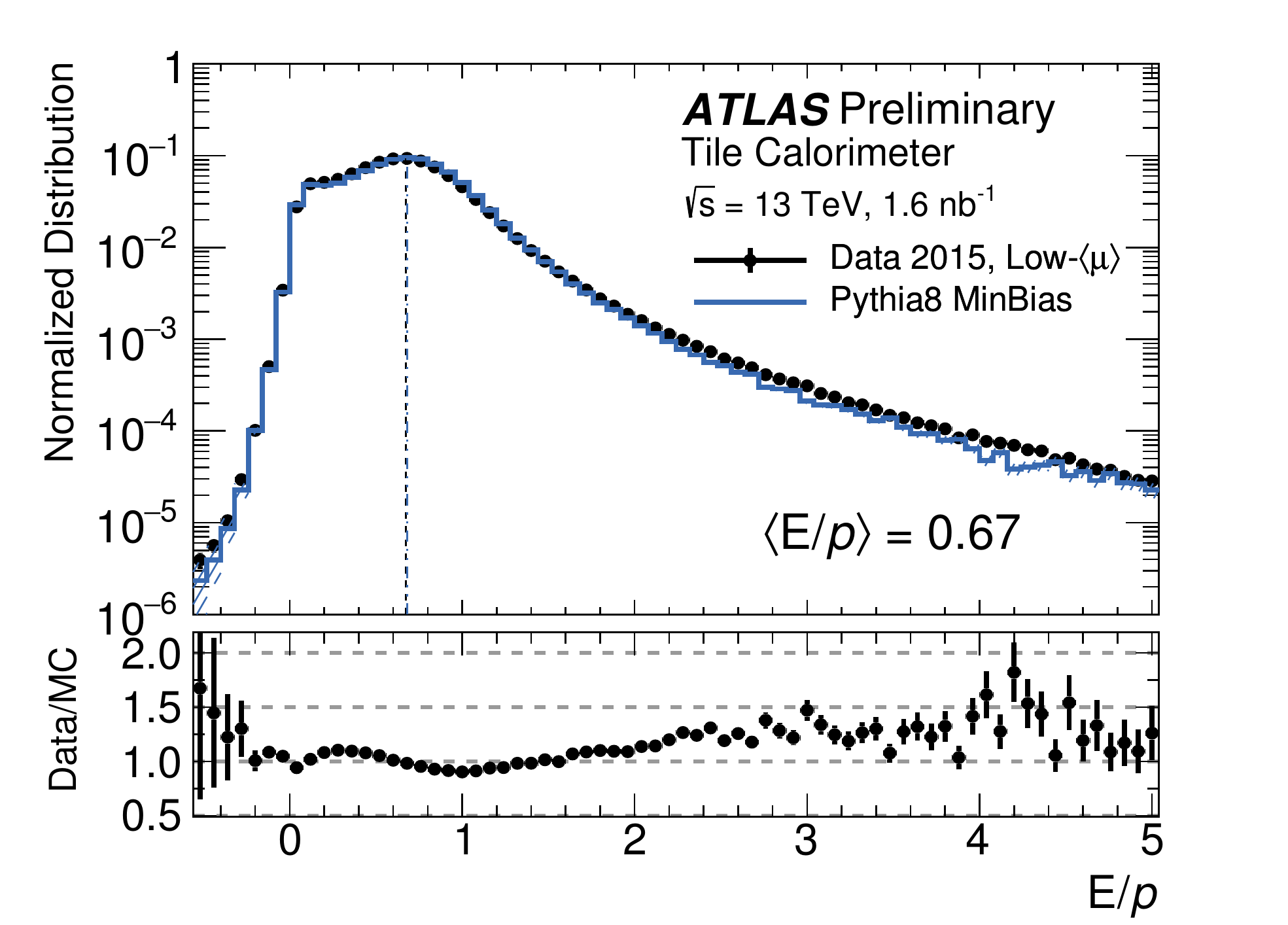}
\caption{Calorimeter response to single isolated charged hadrons.}
\label{fig_isolated_charged_hadrons}
\end{figure}

Figure~\ref{fig_isolated_charged_hadrons} shows the TileCal response to single isolated charged 
hadrons, characterized by energy over momentum ($E/p$), measured using $1.6~{\rm fb^{-1}}$ of 
proton-proton collision data at 13 TeV collected in 2015. The fraction of events with additional 
simultaneous proton-proton collisions (pile-up) was negligible. The data is compared to simulated 
events generated using PYTHIA 8.186. The energy is reconstructed from topological clusters 
matched to a track in a cone of $\Delta R < 0.2$. Tracks are required to pass minimum quality 
criteria and $p > 2~{\rm GeV}$. To reduce contamination from neutral hadrons and muons, the 
energy deposited in the electromagnetic calorimeter is required to be less than $1~{\rm GeV}$, and 
the fraction of energy deposited in the TileCal at least $70\%$. Black dots represent data 
and the blue line represents simulation. Statistical uncertainties are shown for data and simulation. 
The lower plot shows the ratio of data to simulation. Good agreement between data and MC 
prediction is observed.  The average of the $E/p$ distribution is 0.67 (below one), as expected for 
a sampling non-compensating calorimeter.

\begin{figure}[!t]
\centering
\includegraphics[width=3.5in]{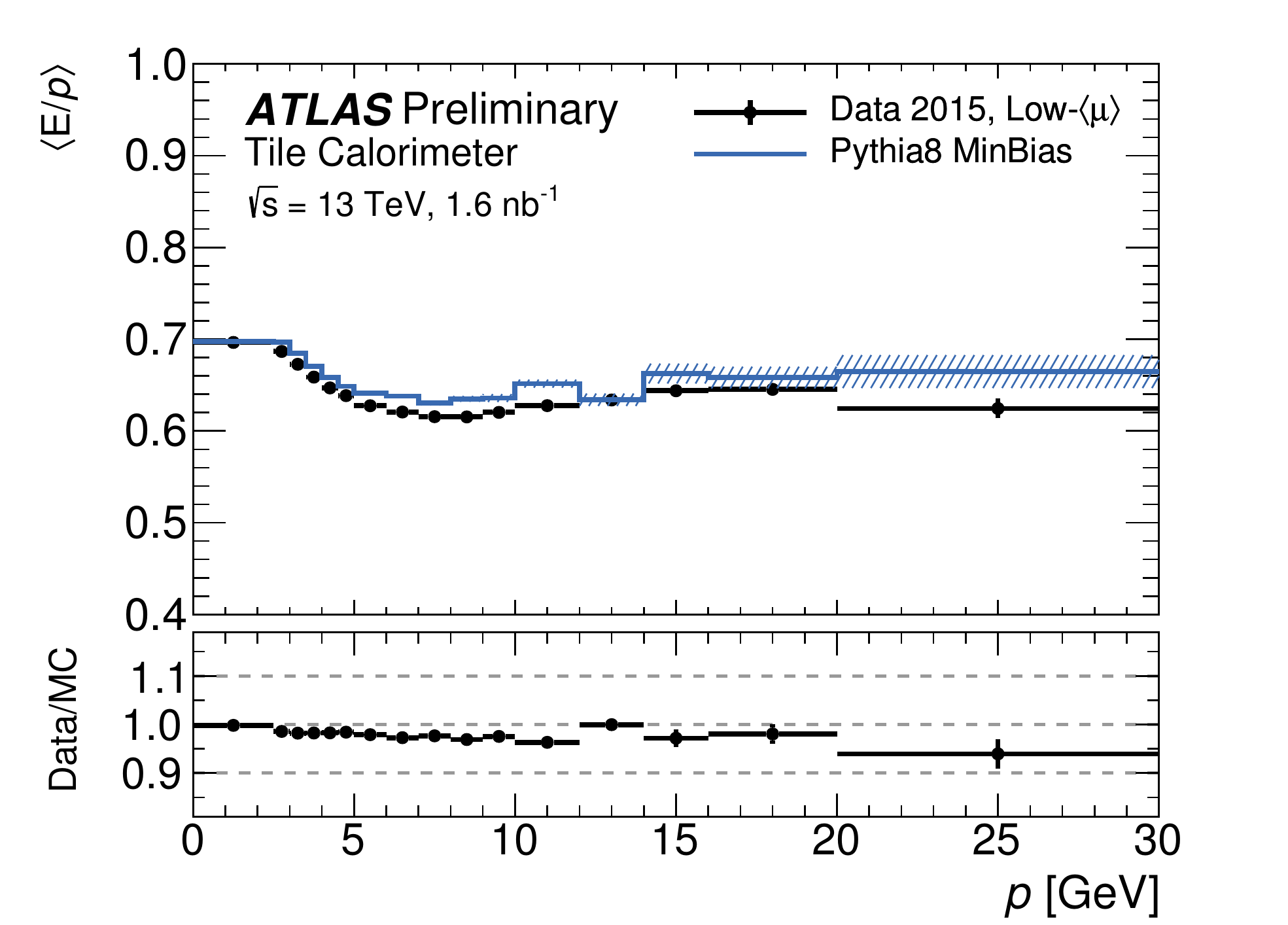}
\caption{TileCal response to single isolated charged hadrons, characterized by the mean of the 
energy over momentum ($\langle E/p\rangle$) as a function of momentum, integrated over the pseudo-rapidity 
and $\phi$ coverage of the calorimeter, measured using $1.6~{\rm fb^{-1}}$ of $pp$ 
collision data at 13 TeV collected in 2015.}
\label{fig_avgeop_vs_p}
\end{figure}

Figure~\ref{fig_avgeop_vs_p} shows the TileCal response to single isolated charged hadrons, characterized by the mean of the energy over momentum ($\langle E/p\rangle$) as a function of momentum, integrated over the pseudo-rapidity and $\phi$ coverage of the calorimeter, measured using $1.6~{\rm fb^{-1}}$ of $pp$ collision data at 13 TeV collected in 2015. The fraction of events with additional simultaneous proton-proton collisions (pile-up) was negligible. The data is compared to simulated events generated using PYTHIA 8.186.

\subsection{Muons}

Muons from cosmic rays are used to study in situ the electromagnetic energy scale and intercalibration of Tile cells.
A good energy response uniformity between calorimeter cells is observed, and 
the response non-uniformity in $\eta$ is measured to be below 5\%.

\begin{figure}[!t]
\centering
\includegraphics[width=3.5in]{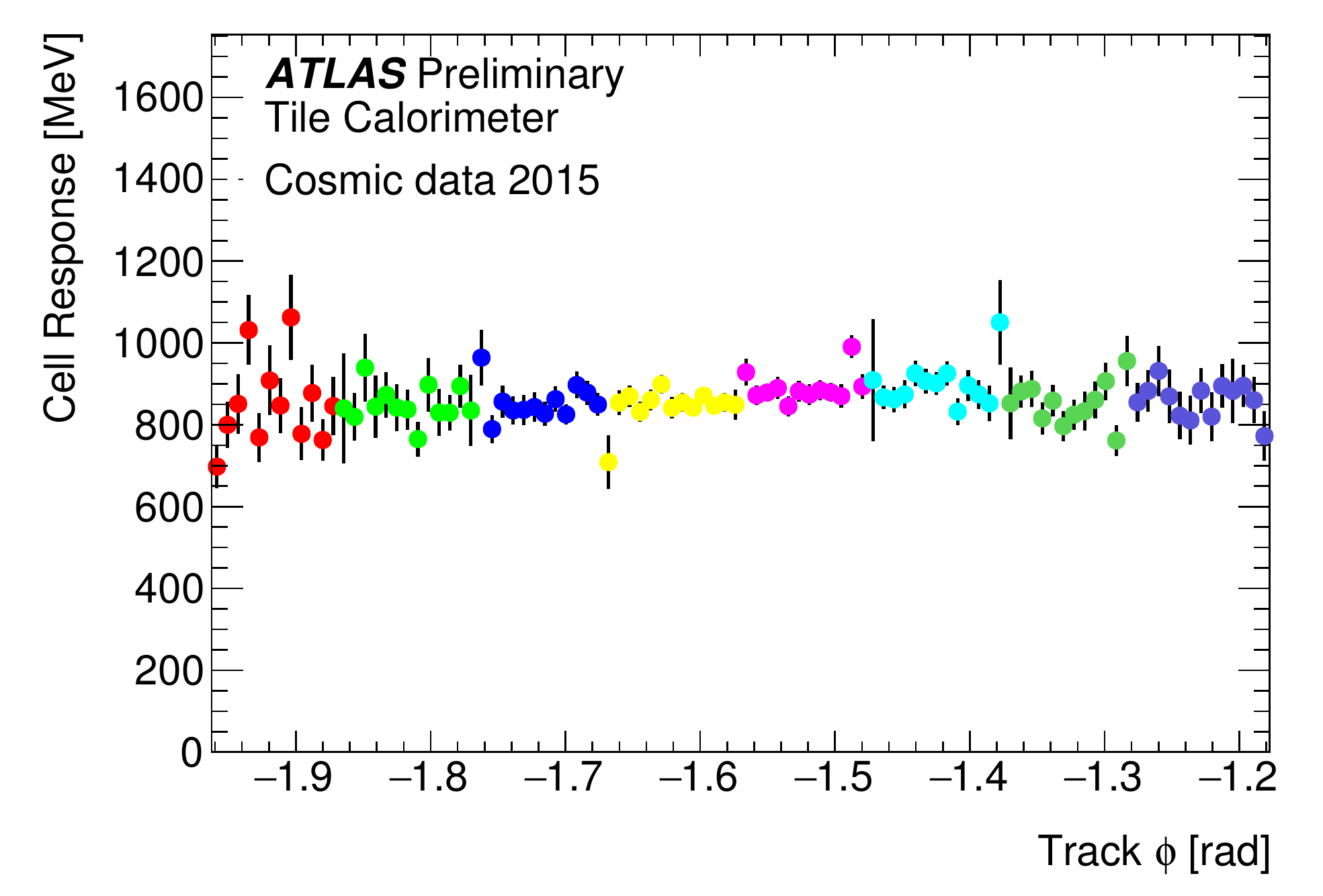}
\caption{Profile of energy deposition in the LB-BC layer as a function of the track $\phi$-coordinate impact point.}
\label{fig_energy_deposition}
\end{figure}

Figure~\ref{fig_energy_deposition} presents the profile of energy deposition in the LB-BC layer as a 
function of the track $\phi$-coordinate impact point, obtained using 2015 cosmic data. The average 
response over all central region cells in each module is shown by symbols of different colors. 
$\phi$ = -1.57 corresponds to vertical track. Only statistical uncertainties are shown.

\subsection{Jet performance}

Good description of the cell energy distribution and of the noise in the calorimeter is crucial for the 
building of topoclusters which are used for jet and missing transverse energy reconstruction

\begin{figure}[!t]
\centering
\includegraphics[width=3.5in]{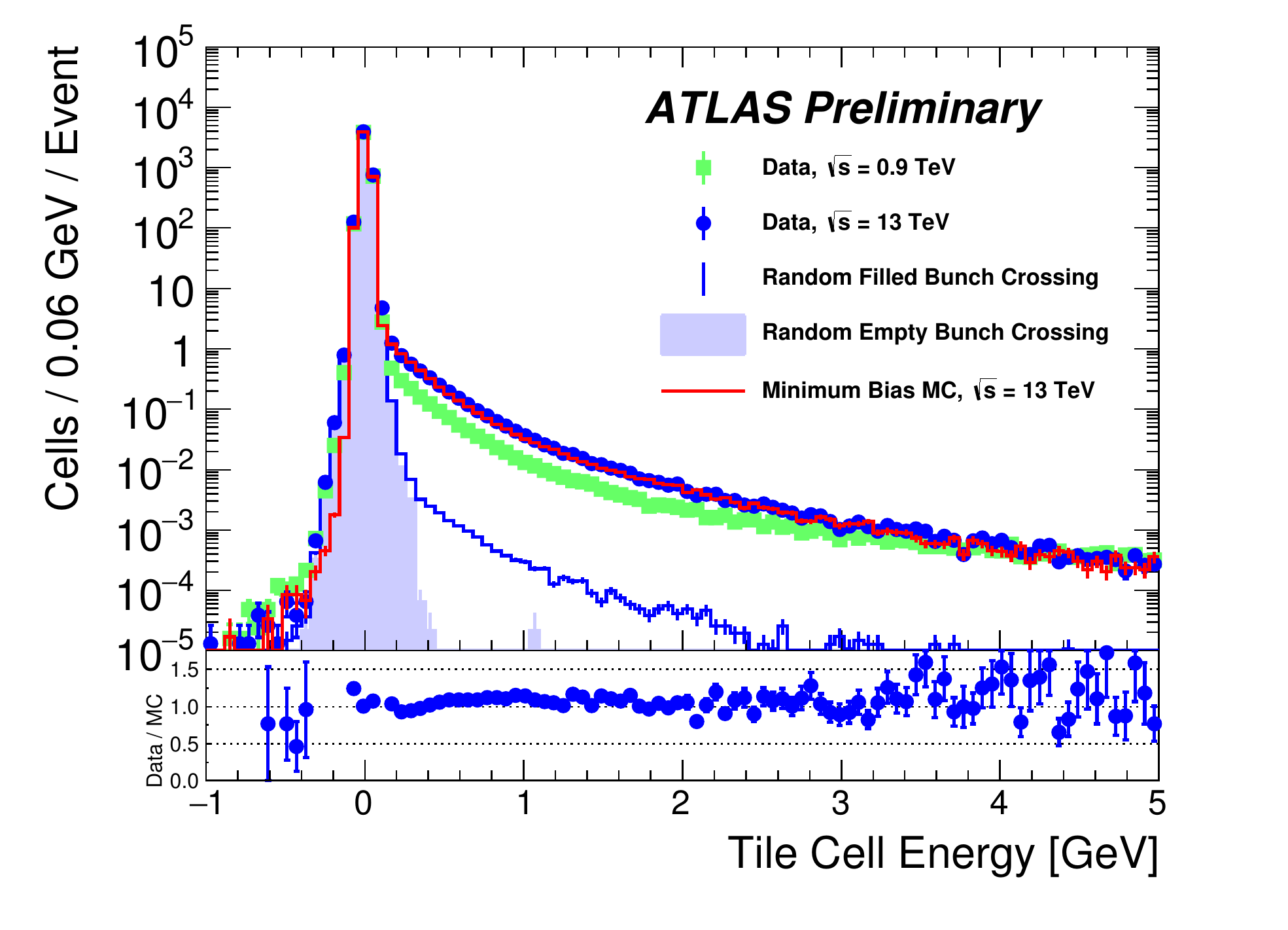}
\caption{Distributions of the energy deposited in the TileCal cells.}
\label{fig_cell_energy_deposition}
\end{figure}

Figure~\ref{fig_cell_energy_deposition} shows the 
distributions of the energy deposited in the TileCal cells from collision data at $\sqrt{s}$ = 13 
and 0.9 TeV superimposed with Pythia minimum bias Monte Carlo and randomly triggered events 
from filled and empty bunch crossings. A Level 1 minimum bias trigger scintillator (MBTS) trigger is 
required for the 13 TeV data. To ensure exactly one interaction has occurred per bunch crossing, 
only events having a single reconstructed primary vertex are selected. Each distribution is 
normalized to unit area. Good agreement is observed in Tile cell energy distribution. 

\begin{figure}[!t]
\centering
\includegraphics[width=3.5in]{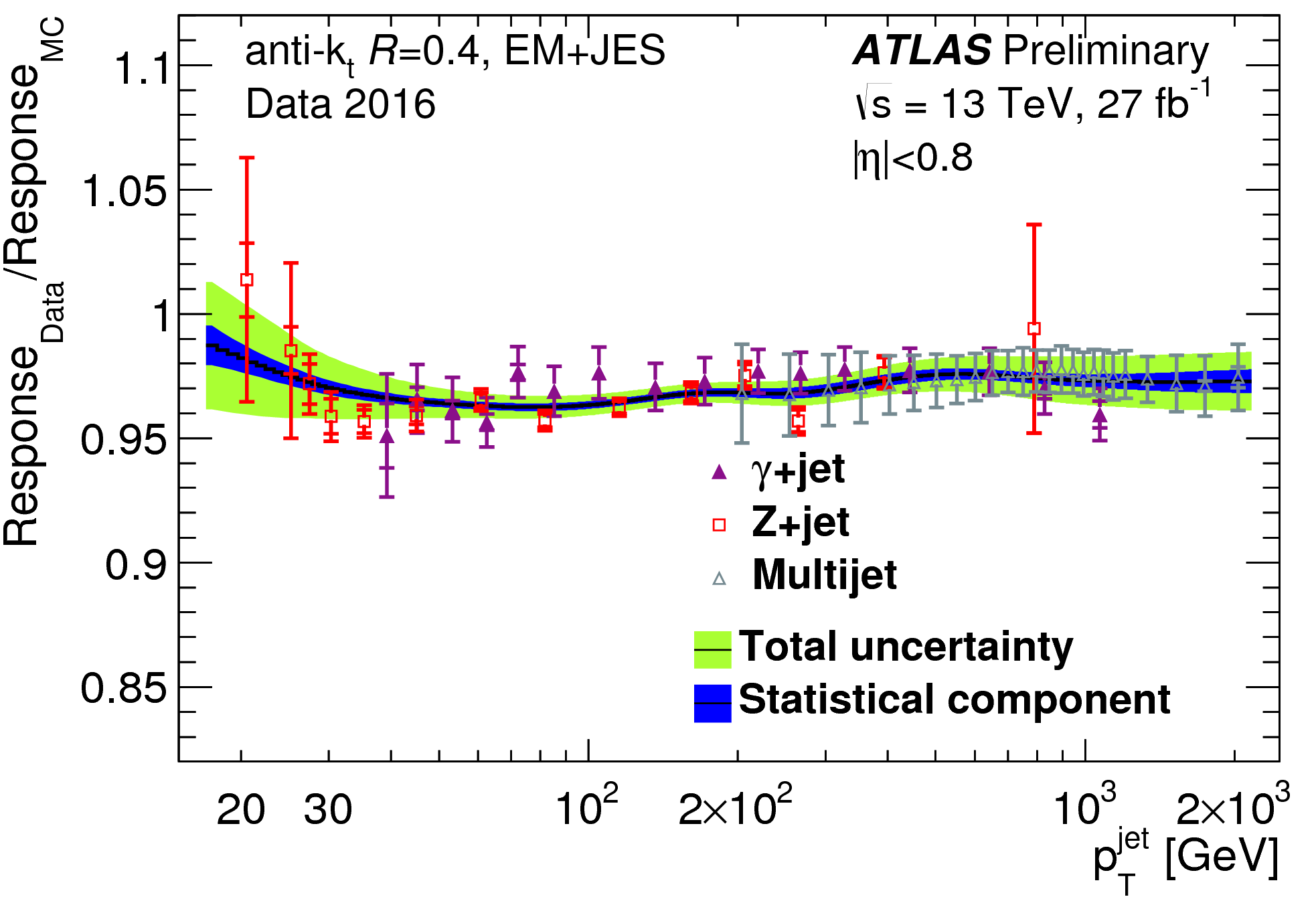}
\caption{Distribution of the jet energy resolution as a function of the $p_T$ of the jet.}
\label{fig_jer}
\end{figure}

Figure~\ref{fig_jer} shows the jet response ratio of the data to the Monte Carlo simulation as a function of $p_T$ for three in situ techniques combined to determine the energy scale correction: Z+jet (squares), $\gamma$+jet (full triangles) and multijet (empty triangles). The error bars showing the statistical and the total uncertainties are presented. The total uncertainty is derived by adding in quadrature statistical and systematic uncertainties. The results are shown for anti-$k_T$ jets with distance parameter of R = 0.4 calibrated with the electromagnetic (EM) and jet energy scale 
(JES) scheme followed by $\eta$ intercalibration and using the 2016 dataset. The result of the combination of the in situ techniques is shown as the dark line. The outer band indicates the total uncertainty resulting from the combination of in situ techniques, while the inner dark band presents the statistical component of the uncertainty. The Z and $\gamma$ analyses are described in~\cite{ATLAS-CONF-2015-057}, the multijet analysis in~\cite{ATLAS-CONF-2015-017}, and the combination in~\cite{ATLAS-CONF-2015-037}. 
Jet energy resolution is around 1\% for $p_T > 100$ GeV and is consistent over all jet energy scale.
Constant term is within expected 3\%.

\section{Conclusion}
The ATLAS Tile Calorimeter is an important sub-detector of the  ATLAS detector at the LHC. 
It is the hadronic sampling calorimeter made of steel plates which act as absorber and scintillating 
tiles as active medium. Control of its energy 
is essential to measure the energy of jets, hadronically decaying tau leptons and missing transverse 
energy.
The TileCal calibration system consists of Cesium radioactive sources, laser, charge injection 
components, and an integrator based readout system. Combined information from all systems allows 
for an efficient monitoring and correction of fine instabilities of TileCal cells response. 
Intercalibration and uniformity are monitored with isolated charged hadrons and cosmic muons. 
Data quality in physics runs is monitored extensively and continuously. All problems are reported 
and addressed. The data quality efficiency achieved was 99.6\% in 2012, 100\% in 
2015, 98.9\% in 2016 and 99.4\% in 2017.
The stability of the absolute energy scale at the cell level was maintained to better than 1\% during 
LHC data-taking.
Following the experience gained during LHC Run-I, all calibration systems were improved for Run-II. TileCal 
performance during LHC Run-II, (2015-2018), including calibration, stability, absolute energy scale, 
uniformity and time resolution show that the TileCal performance is within the design requirements 
and has given essential contribution to reconstructed objects and physics results.


%





\ifCLASSOPTIONcaptionsoff
  \newpage
\fi

\end{document}